\documentclass[conference,10pt]{IEEEtran}

\usepackage{bm, latexsym, amsmath, amssymb, times}
\usepackage{cite, setspace, authblk, color}
\usepackage{rotating, multirow, booktabs, psfrag, epsfig, wrapfig}
\usepackage[caption=false, font=footnotesize]{subfig}
\usepackage{theorem, verbatim, algpseudocode, algorithmicx, algorithm}
\usepackage[left=1.62cm,right=1.62cm,top=1.9cm]{geometry}

\IEEEoverridecommandlockouts
\allowdisplaybreaks

\begin{document} 
        \title{Adversarial Attacks and Defenses for Wireless Signal Classifiers using CDI-aware GANs}

        \author{Sujata Sinha}
        \author{Alkan Soysal}
        \affil{\normalsize Wireless@VT, Bradley Department of Electrical and Computer Engineering Virginia Tech, Blacksburg, VA 24061} 
        \maketitle
\vspace{-.3in}

\begin{abstract}
We introduce a Channel Distribution Information (CDI)-aware Generative Adversarial Network (GAN), designed to address the unique challenges of adversarial attacks in wireless communication systems. The generator in this CDI-aware GAN maps random input noise to the feature space, generating perturbations intended to deceive a target modulation classifier. Its discriminators play a dual role: one enforces that the perturbations follow a Gaussian distribution, making them indistinguishable from Gaussian noise, while the other ensures these perturbations account for realistic channel effects and resemble no-channel perturbations.

Our proposed CDI-aware GAN can be used as an attacker and a defender. In attack scenarios, the CDI-aware GAN demonstrates its prowess by generating robust adversarial perturbations that effectively deceive the target classifier, outperforming known methods. Furthermore, CDI-aware GAN as a defender significantly improves the target classifier’s resilience against adversarial attacks.
\end{abstract}

\section{Introduction}
\label{sec:intro}
DNN models, despite their adaptability to data, are vulnerable to adversarial attacks. Adversaries exploit the blind spots of these models, undermining their strengths. Adversaries evade a trained target classifier by exploiting system vulnerabilities to craft perturbations that deceive the target system~\cite{biggio2013evasion, goodfellow2014explaining}. 

Vulnerabilities to adversarial attacks extend to any system that employs a DNN, including wireless communications~\cite{adesina2022adversarial, Sinha2023, flowers2019evaluating, lin2020threats,  sadeghi2018adversarial, kim2021channel, bahramali2021robust, Hameed2021}. For example, autoencoder-based communication systems and signal detection in OFDM are susceptible to such attacks~\cite{bahramali2021robust}. AI techniques are also commonly employed to disrupt radio access in 5G and 6G systems~\cite{shi2022attack, shi2021adversarial, siriwardhana2021ai}, as well as in cooperative spectrum sensing~\cite{luo2020attackers}. 

Automatic modulation classifiers use DNN models to categorize incoming signals into distinct modulation classes. An over-the-air adversary exploits the properties of DNNs to craft adversarial perturbations that fool the receiver into making incorrect class predictions. Attackers generate perturbations through methods that are either input-dependent~\cite{sadeghi2018adversarial, lin2020threats, flowers2019evaluating, kim2021channel} or input-agnostic~\cite{sadeghi2018adversarial, lin2020threats, Hameed2021, bahramali2021robust, kim2021channel}. 

Universal adversarial perturbations (UAPs), employing an input-agnostic approach, aim to deceive the DNN model at the legitimate receiver using a single perturbation vector. However, these single-vector UAP attacks are susceptible to easy detection and mitigation by the receiver. To overcome this limitation,~\cite{bahramali2021robust} propose an alternative using a GAN to craft a new perturbation signal for each channel use.

Both traditional single-vector UAPs~\cite{sadeghi2018adversarial} and the newer GAN-based perturbations~\cite{bahramali2021robust} have been developed under ``white-box" or ``black-box" attack configurations. In ``white-box" attacks, the adversary is assumed to have complete knowledge of the target DNN model and its parameters. Conversely, ``black-box" attacks presume that the adversary has no information about the target DNN classifier. However, these classifications are not fully applicable to wireless communication systems.

Unlike computer vision applications, a wireless adversary cannot directly manipulate data at the input to the classifier. The perturbations experience phase and amplitude changes as they propagate through the channel to reach the legitimate receiver. The effectiveness of these received perturbations depends on the extent of the adversary's system knowledge and their ability to compensate for realistic channel effects. One attack model considered in the literature assumes that the attacker knows the unattacked received signal, the instantaneous channel between the attacker and receiver, the training dataset, and the target classifier model \cite{flowers2019evaluating, lin2020threats, Hameed2021, sadeghi2018adversarial, kim2021channel}. Another model assumes the instantaneous channel knowledge, the training dataset, and the target classifier model for crafting input-agnostic UAP perturbations \cite{sadeghi2018adversarial, kim2021channel}. However, ``black-box attacks", as described in \cite{sadeghi2018adversarial, lin2020threats, Hameed2021, kim2021channel, bahramali2021robust}, are ambiguously labeled, as they either presume channel knowledge or disregard channel effects entirely. These models lack practicality in over-the-air communications, where exact channel responses are unknown to the attacker. In this paper, we consider a model where the attacker is only aware of the statistical properties of channel coefficients and the training dataset, similar to ``channel-independent UAP attacks" in~\cite{kim2021channel}. We propose GAN-based attacks and defenses that outperform those in~\cite{kim2021channel} when the attacker only knows the distribution of the channel.

\emph{Channel-aware} attacks are designed to consider realistic channel effects, ensuring the deception of the legitimate receiver despite changes in amplitude or phase of the transmitted perturbation during signal transmission. Kim et al.~\cite{kim2021channel} propose varied strategies for creating these perturbations, each adapted to different degrees of information uncertainty during the training phase. However, \emph{channel-aware} UAPs crafted using their methods provide a singular attack vector, making the attack easier to detect.

In this paper, we propose an input-agnostic CDI-aware GAN. Instead of crafting a single UAP vector, this CDI-aware GAN learns a distribution over channel-aware perturbations, offering the user a new perturbation signal for each channel use. Additionally, our CDI-aware GAN aligns the perturbations with the expected additive white Gaussian noise profile of wireless channels while also embedding the adversarial characteristics of the Fast Gradient Method (FGM) based perturbations. Consequently, the generated perturbations not only mislead the legitimate classifier but also blend in as channel noise.

Our CDI-aware GAN method is applicable both as a tool for wireless adversaries and defenders. Initially, we analyze the performance of the CDI-aware GAN as an attacker against the classifier at the legitimate receiver without any defensive countermeasures in place. Subsequently, we explore the CDI-aware GAN's defensive capabilities. Our results reveal that in the role of an attack generator, our model surpasses existing state-of-the-art channel-aware attack strategies by 3 dB. More notably, the CDI-aware GAN significantly excels in its defensive role. We detail its effectiveness in countering channel-aware gradient-based FGM attacks, optimization-based Perturbation Generator Model (PGM) attacks, and attacks generated by our own CDI-aware model. We conclude that, as a defense mechanism, the CDI-aware GAN notably enhances the robustness of the legitimate classifier against over-the-air adversarial attacks.

\section{System Model} 
\label{sec: System Model}
We consider a wireless communication scenario with a legitimate transmitter/receiver pair and an adversary. The legitimate receiver employs a DNN-based modulation classifier to learn the modulation type the transmitter utilized. When the transmitted waveform is $x(t)$, and in the absence of an attacker, the received waveform passing through a slowly fading and frequency flat wireless channel is given as
\begin{align}
    r(t) = h_{rt} x(t) + n(t),
\end{align}
where $h_{rt}$ is the channel coefficient from the transmitter to the receiver and $n(t)$ is additive white Gaussian noise. In order to use $r(t)$ as an input to an ML model, the receiver needs a feature extraction process that can be modeled as a function $g[\cdot]$ from the waveform space to the feature space. Let us denote the feature vector as $\mathbf{x}_{e} = g[r(t)]$. Training of an ML model by such a dataset adds another layer of transformation, $f(\cdot, \bm{\theta})$, that provides the class decisions given the features of the signal, $\mathbf{x}_{e}$. The overall transformation from the waveform to the class decisions is given by
\begin{align}
    f \left( \mathbf{x}_{e}, \bm{\theta} \right) = f \left( g \left[ h_{rt} x(t) + n(t)\right] , \bm{\theta} \right).
\end{align}
\begin{figure}[t]
    \centering
    \subfloat[]{\includegraphics[width=\columnwidth]{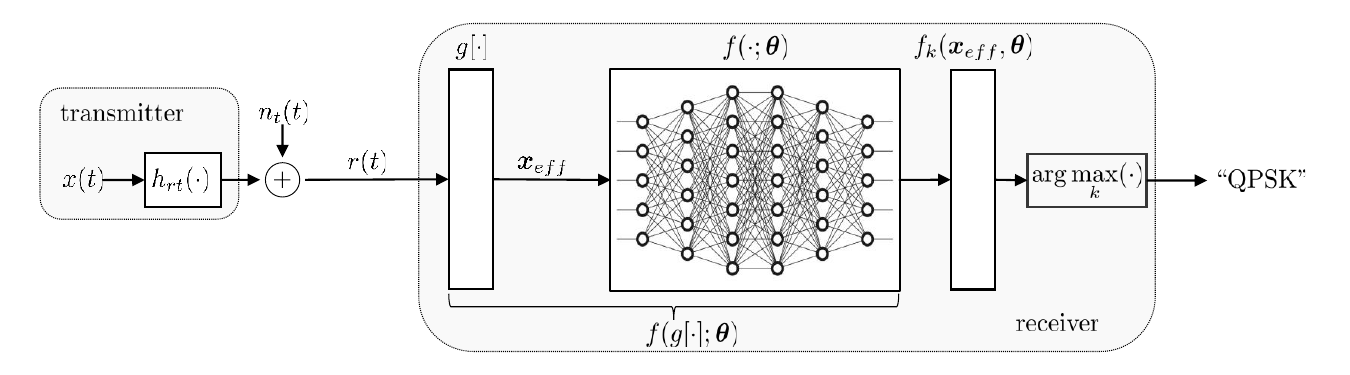}} \\
    \subfloat[]{\includegraphics[width=\columnwidth]{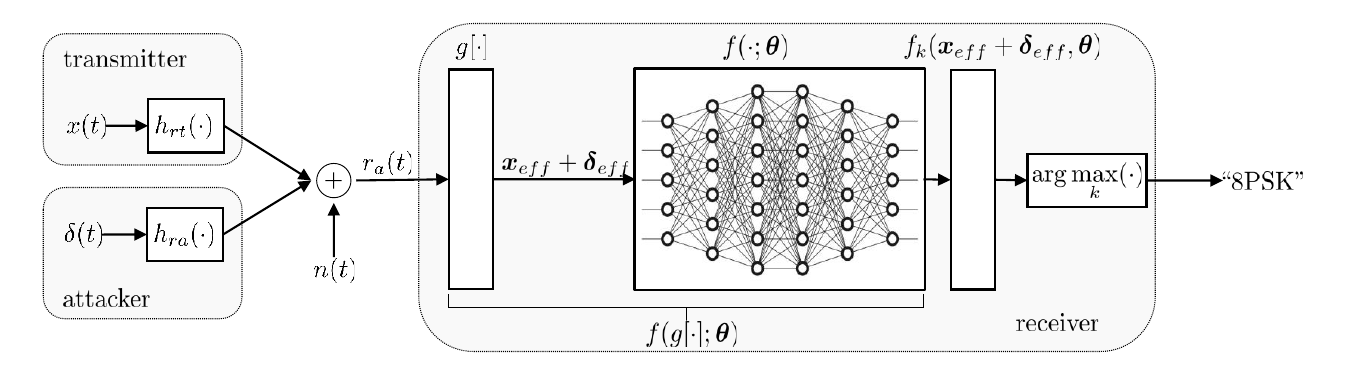}}
    \caption{An over-the-air modulation classifier: (a) a QPSK waveform $x(t)$ is correctly classified; (b) a QPSK signal $x(t)$ is attacked and misclassified as an 8PSK signal.}
    \label{fig:ModClass}
\end{figure}

We can regard the composite function, $f(g[\cdot], \bm{\theta} )$, as the ``effective model''. ML-based signal classification literature typically chooses a $g[\cdot]$ as high-dimensional I/Q time samples of the received waveform \cite{o2016radio}, creates a dataset using $g[\cdot]$, chooses a DNN architecture, and finally trains a model, $f(\cdot, \bm{\theta})$, to decide on the class (e.g., modulation) of an incoming waveform \cite{o2016radio, o2017introduction}. Note that the input to the classifier is $\mathbf{x}_{e} = g[r(t)] \in \mathcal{X} \subset \mathbb{C}^{p}$, and the associated true labels are $\bm{\ell}_{true}$, where $p$ is the number of complex-valued samples in the dataset. Fig.~\ref{fig:ModClass}(a) illustrates this process for a modulation classifier, which is trained using $\mathbf{x}_{e}$, I/Q time samples of the received waveform.

An over-the-air adversary attempts to introduce a perturbation waveform, $\delta(t)$, to fail the target classifier at the legitimate receiver. However, the adversarial attack to the modulation classifier suffers from the effects of over-the-air transmission, and the received signal at the receiver becomes 
\begin{align}
    r_\delta(t) &= h_{rt} x(t)  + h_{ra} \delta(t)  + n(t) \\
            &= r(t) + h_{ra} \delta(t) , \label{eqn:advWaveform}
\end{align}
where $h_{ra}$ is the channel coefficient from the attacker to the receiver. In the case where I/Q time samples are used, (\ref{eqn:advWaveform}) becomes%
\begin{align}
    \mathbf{x}_{adv} &= \mathbf{x}_{e} + \bm{\delta}_{e} \\
    &= \mathbf{x}_{e} + \mathbf{H}_{ra}\bm{\delta},
    \label{eqn: xadv}
\end{align}%
where $\mathbf{H}_{ra} = \mbox{diag}\{ {h}_{ra_1}, \ldots,h_{ra_{p}} \} \in \mathbb{C}^{p \times p} $ is the diagonal channel matrix between the attacker and receiver. Fig.~\ref{fig:ModClass}(b) illustrates this process, where the modulation classifier observes I/Q time samples of the adversarial waveform, $\mathbf{x}_{adv}$. 

Within this framework, the adversary crafts the perturbation, $\bm{\delta}_e$, by formulating the objective as follows 
\begin{align}
& \arg \min_{\bm{\delta}_{e}} \Vert \bm{\delta}_{e} \Vert_{2} \label{eq: armindelta} \\
& \text{s.t. } \forall \mathbf{x}_{e} \in \mathcal{X} : f(\mathbf{x}_{adv}, \bm{\theta}) \neq f(\mathbf{x}_{e}, \bm{\theta}) \label{eq: condition1} \\
&   \Vert \bm{\delta}_{e} \Vert_{2}^{2} \leq p_{max} \label{eq: condition2}
\end{align}
where $\Vert \cdot \Vert_{2}$ represents $l_2$ norm and restricts the power budget of the crafted perturbations to $p_{max}$. The goal of the adversary is to find the universal adversarial perturbation vector $\bm{\delta}_{e}$ under two constraints. The first constraint in (\ref{eq: condition1}) requires the adversarial signal to cause misclassification at model $f(\cdot, \bm{\theta})$. The second constraint in (\ref{eq: condition2}) ensures that the attack on $f(\cdot, \bm{\theta})$ is covert. 

Finally, to measure the effectiveness of adversarial perturbations, we adopt perturbation-to-noise ratio (PNR) as a metric\cite{sadeghi2018adversarial},
\begin{align}
\label{eqn:pnr}
    \text{PNR (dB)} &= \frac{\text{Received Perturbation Power ($\mathrm{P_{rx}}$)}}{\text{Noise Power ($\mathrm{P_{\mathbf{n}}}$)}} \text{ (dB)} 
\end{align}

The success of an adversarial attack is largely contingent upon the attacker's knowledge of the system. In our study, we assume the attacker is aware of the statistical properties of the channel coefficients from the attacker to the receiver, the training dataset, and the target modulation classifier, denoted as $f(\cdot, \bm{\theta})$. The channel in this context is modeled as a composite of Rayleigh distributed fast fading, lognormal distributed shadowing, and deterministic path loss.

\section{CDI-aware GAN} \label{sec: CDI_Aware_Generator_Model}

\begin{figure}
    \centering\includegraphics[width=\linewidth]{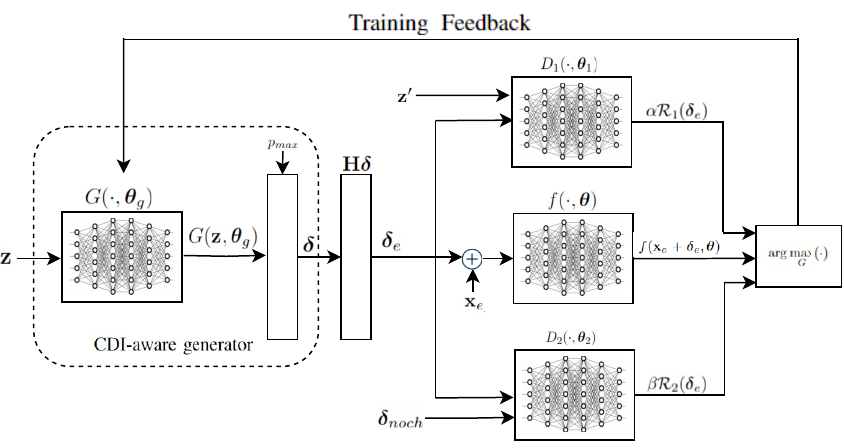}
    \caption{Training of CDI-aware GAN with a generator ($G$), two discriminators and the underlying DNN classifier. The discriminators ($D_1$) and ($D_2$) are trained simultaneously with generator $G$ and ensure that the received perturbations are indistinguishable from AWGN and offer model fooling capabilities of \emph{no-channel} perturbations, respectively.}
    \label{fig: CDI_aware_generator_model_training}
\end{figure}

GANs operate by training two models simultaneously: a generative model, which learns to mimic the data distribution, and a discriminative model, which discerns whether a sample originates from the actual data or the generative model. This iterative training process is designed to enhance the generative model until it produces data nearly indistinguishable from real-world data.

In our work, we present a CDI-aware GAN that crafts input-agnostic perturbation signals from random input noise, $\mathbf{z}$, with the objective of maximizing the loss of the modulation classifier. The generator in our model, designated as $G$ and parametrized by $\bm{\theta}_g$, is a differentiable function that learns distribution $p_{\delta}$ over perturbations $\bm{\delta}$. These crafted perturbations are transmitted over the adversary-receiver channel, $\mathbf{H}$, resulting in $\bm{\delta}_e$. The discriminator components, $D_{1}$ and $D_2$, with parameters $\bm{\theta}_{1}$ and $\bm{\theta}_{2}$ respectively, are binary classifiers. They are simultaneously trained with generator $G$ to accurately identify additive white Gaussian noise and no-channel perturbations $\bm{\delta}_{noch}$ from perturbation $\bm{\delta}_e$. Fig.~\ref{fig: CDI_aware_generator_model_training} illustrates our CDI-aware GAN model.  

\subsection{Training}
Given a random input noise variable $\mathbf{z}$ drawn from a distribution $p_{\mathbf{z}}(\mathbf{z})$, our generator functionally maps $\mathbf{z}$ to the feature space $G(\mathbf{z}, \bm{\theta}_g)$. This approach allows us to move beyond creating a single UAP; instead, we learn a distribution $p_{\delta}$ from which we can sample multiple perturbations. These sampled perturbations, denoted as $\bm{\delta} \sim p_{\delta}$, are subject to a power constraint, $p_{max}$. To comply with this constraint for each perturbation $\delta = G(\mathbf{z}, \bm{\theta}_g)$, similar to~\cite{bahramali2021robust}, we implement a remapping strategy as,
\begin{align}
    \boldsymbol{\delta}
    = & \begin{cases} 
     G(\mathbf{z}, \bm{\theta}_{g}) & \text{if } \left\|G(\mathbf{z}, \bm{\theta}_{g})\right\|_{2}^{2} \leq p_{max} \\
     \sqrt{p_{max}}\frac{G(\mathbf{z}, \bm{\theta}_{g})}{\|G(\mathbf{z}, \bm{\theta}_{g})\|_{2}} & \text{otherwise}
    \end{cases}
\end{align}
Assuming the channel between the adversary and the legitimate receiver follows the distribution $p_{ra}$, the channel realization matrix can be represented as $\mathbf{H} = \mbox{diag}\{ {h}_{1}, \ldots,h_{{p}} \}$, where each matrix element $h_i$ is drawn from $p_{ra}$. Consequently, the received perturbation in I/Q time samples, $\bm{\delta}_{e}$, is given by $\mathbf{H}\boldsymbol{\delta}$. 

We define the optimization problem as 
\begin{align}
        \arg\max_{G} \displaystyle \mathop{\mathbb{E}}_{\mathbf{z} \sim p_{\mathbf{z}}(\mathbf{z}),\text{ } h_i \sim p_{ra}} \left[\sum_{\mathbf{x}_{e} \in \mathcal{X}} L(\bm{\ell}_{adv}, \bm{\ell}_{pred} )\right]
    \label{eq: pgm_optimization}
\end{align}
where $\bm{\ell}_{adv}$ and $\bm{\ell}_{pred}$ are the prediction of the legitimate classifier on $\mathbf{x}_{adv}$ and $\mathbf{x}_{e}$, respectively and $L$ is the classifier's loss function. In this setting, our goal is to maximize the expected value of the cumulative loss for the legitimate classifier. This approach ensures that the perturbations, even after traversing the adversary-receiver channel, effectively mislead the DNN model.

Under the training criterion outlined in~(\ref{eq: pgm_optimization}), the generator model is trained to create optimized vectors within the feature space $\mathbf{x}_{e}\in \mathcal{X}$. As a result, the crafted perturbations acquire unique statistical properties, making them distinct from the time samples of additive white Gaussian noise. Discriminator $D_{1}$ enforces a Gaussian distribution on the signal $\mathbf{H}\bm{\delta}$, making it indistinguishable from $\mathbf{n}$.

Furthermore, the perturbation $\bm{\delta}$ is designed to counteract the effects of the channel $\mathbf{H}_{ra}$. The optimal solution generates perturbations so that~(\ref{eqn: xadv}) simplifies to $\mathbf{x}_{adv} = \mathbf{x}_{e} + \bm{\delta}_{noch}$. Discriminator $D_{2}$ ensures that the created perturbation signal effectively inverts the statistical effects of the channel, leading to an observed signal $\bm{\delta}_{e}$ that closely resembles $\bm{\delta}_{noch}$. 

Our adversarial modeling framework trains the generator and discriminators simultaneously by introducing a regularization effect in the optimization strategy of~(\ref{eq: pgm_optimization}) and modifying it as
\begin{align}
\hspace{-5pt}
\arg\max_{G} \displaystyle \mathop{\mathbb{E}}_{\mathbf{z},\text{ } \mathbf{H}} \left[ \sum_{\mathbf{x}_{e} \in \mathcal{X}} \!L(\bm{\ell}_{adv}, \bm{\ell}_{pred} )  \!+ \!\alpha\mathcal{R}_{1}(\bm{\delta}_e) \!+ \!\beta\mathcal{R}_{2}(\bm{\delta}_e)\right]
\hspace{-5pt}
\label{eq: final_optimizer}
\end{align}
where $\alpha$ and $\beta$ are weights of the regularizer relative to the objective function described in~(\ref{eq: pgm_optimization}). Discriminators $D_1$ and $D_2$ enforce regularization $\mathcal{R}_{1}$ and $\mathcal{R}_{2}$, respectively. The generator $G$ assisted by regularizers $D_1$ and $D_2$ implicitly define distribution $p_{\delta}$ that converges to a cumulative distribution function that shares characteristics with $\mathcal{N}(\mu, \sigma^2)$ and $\bm{\delta}_{noch}$. Algorithm~\ref{algo: training_pgm} summarizes our technique to train the CDI-aware GAN.

\begin{algorithm}[t]
    \caption{CDI-aware GAN Training}
    \begin{algorithmic}[1]
    \State \textbf{Inputs: }$\mathcal{X}$, $f(\cdot, \bm{\theta})$, $L$, $p_{max}$, $G(\cdot, \bm{\theta}_g)$, $D_{1}(\cdot, \bm{\theta}_1)$, $D_{2}(\cdot, \bm{\theta}_2)$, $p_{ra}$, and $K$
    \For{epoch $k \gets 1$ to $K$}
        \For{mini-batch $b$ in $\mathcal{X}$}
            \State Sample $\mathbf{z} \sim p_{\mathbf{z}}(\mathbf{z})$, $\mathbf{z'} \sim \mathcal{N}(\mu, \sigma^2)$ and $\mathbf{H}$
            \State Compute $\bm{\delta}_{noch} = \sqrt{p_{max}}\frac{\nabla_{\mathbf{x}} L(f(\mathbf{b}, \bm{\theta}))}{\Vert \nabla_{\mathbf{x}} L(f(\mathbf{b}, \bm{\theta}))\Vert_{2}}$
            \State Generate $G(\mathbf{z}, \bm{\theta}_g)$ 
            \State Enforce constraint $p_{max}$ on $G(\mathbf{z}, \bm{\theta}_g)$ to obtain $\bm{\delta}$
            \State Compute signal $\bm{\delta}_{e}$ = $\mathbf{H}\bm{\delta}$
            \State Train $D_1$ on $\bm{\delta}_{e}$ (label $1$) and $\mathbf{z'}$ (label $0$)
            \State Train $D_2$ on $\bm{\delta}_{e}$ (label $1$) and $\bm{\delta}_{noch}$ (label $0$)
            \State Minimize the objective function~(\ref{eq: final_optimizer})
            \State Update $G$ to minimize loss $L$
        \EndFor
    \EndFor
    \end{algorithmic}
    \label{algo: training_pgm}
\end{algorithm}

The CDI-aware GAN is versatile, finding utility both as a tool for wireless adversaries and defenders. In an offensive role, an over-the-air adversary utilizes Algorithm~\ref{algo: training_pgm} to train the CDI-aware GAN during the training phase. This preparation allows the adversary to effectively use the trained GAN model in the testing phase for creating and transmitting adversarial perturbation signals toward the legitimate receiver. 

In an offensive role, the legitimate receiver employs adversarial training to enhance the robustness of the classifier model. This process involves the defender generating an adversarial perturbation signal and integrating it into the training dataset. The classifier is then retrained with this augmented dataset, which comprises both the original In-phase and Quadrature (I/Q) time samples and the newly crafted adversarial examples. Through this iterative training approach, the classifier not only becomes resilient to adversarial attacks but also retains its proficiency in accurately classifying legitimate inputs.

\section{CDI-Aware GAN as an Attacker}
\begin{figure}[t]
    \centering
    {\includegraphics[width=\linewidth]{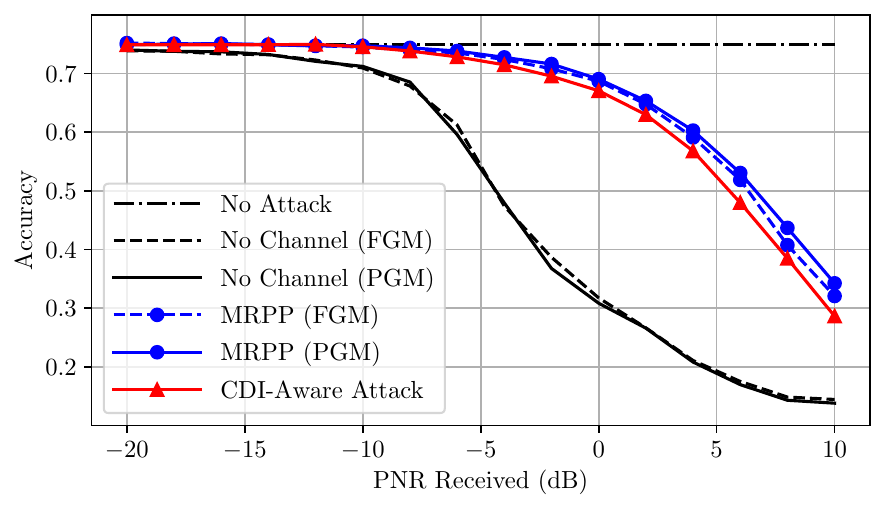}}
    \caption{Our adversarial attack, based on CDI-aware GAN, compared to other known attacks. We assume the target model does not use any defense strategy. }
    \label{fig: attacks}
    \vspace{-12pt}
\end{figure}

In this section, we present the performance evaluation of our CDI-aware GAN in the role of an attacker, alongside a comparative analysis with various attacks documented in the literature. Figure~\ref{fig: attacks} demonstrates the impact of different attack strategies on the accuracy of the classifier at the legitimate receiver, depicted as a function of received PNR. In scenarios devoid of any attack, the classifier maintains a constant level of accuracy. In this \emph{no-channel} attack scenario, the channel matrix is assumed to be an identity matrix, $\mathbf{H}_{ra} = \mathbf{I}$. This scenario represents an ideal case for the attacker, as the absence of channel effects means there is no distortion of the designed perturbation signal, typically resulting in the lowest classifier accuracy as depicted in Figure~\ref{fig: attacks}.

In our comparative analysis, we evaluate our CDI-aware GAN attack against established methods such as the UAP-FGM attack from~\cite{sadeghi2018adversarial} and the PGM attack by~\cite{bahramali2021robust}, both designed for no-channel scenarios. The UAP-FGM approach generates a single input-agnostic perturbation from a randomly selected sample set of training data, whereas the PGM method produces unique perturbations for each channel use, based on triggers from a uniform distribution ($\sim \text{Uniform}(0,1)$). 

Additionally, we incorporate the Maximum Received Perturbation Power (MRPP) attack model from~\cite{kim2021channel}, which considers channel effects. This model applies an MRPP transformation to the UAP-FGM attack of \cite{sadeghi2018adversarial} when the attacker has either perfect or CDI-only channel knowledge. For the CDI-aware case that we are interested in, the attacker of \cite{kim2021channel} first crafts a perturbation signal using UAP-FGM, then calculates the MRPP transformation using a sample set of channel realizations drawn from the channel distribution. This attack is denoted as MRPP (FGM) in Fig.~\ref{fig: attacks}. 

To ensure a balanced comparison, we introduce a novel attack, denoted as MRPP (PGM) in Fig.~\ref{fig: attacks}, applying the MRPP transformation to the PGM attack of \cite{bahramali2021robust}, akin to the methodology in \cite{kim2021channel}. 

Our proposed CDI-Aware GAN attack differs from the MRPP with PGM approach, as it integrates CDI directly into the GAN architecture rather than relying on a limited subset of channel realizations like MRPP. Furthermore, our method crafts perturbation signals for transmission without the need for extra mathematical operations, and like PGM, it generates a new perturbation for each channel use.

Figure~\ref{fig: attacks} demonstrates the superior effectiveness of our proposed CDI-aware attack compared to the FGM-based and PGM-based MRPP attacks. The PGM-based MRPP attack shows similar performance to its FGM-based counterpart. This is consistent with \cite{bahramali2021robust}, where there is no fading channel. The advantage of PGM-based attacks is their undetectability. On the other hand, our CDI-aware attack outperforms both, requiring approximately 3 dB less PNR to achieve the same level of accuracy. This finding is significant given the critical importance of detectability in wireless adversarial contexts. The subsequent section will further illustrate the prowess of our CDI-aware GAN approach, particularly as a defender.

\begin{figure}[t]
    \centering
    {\includegraphics[width=\linewidth]{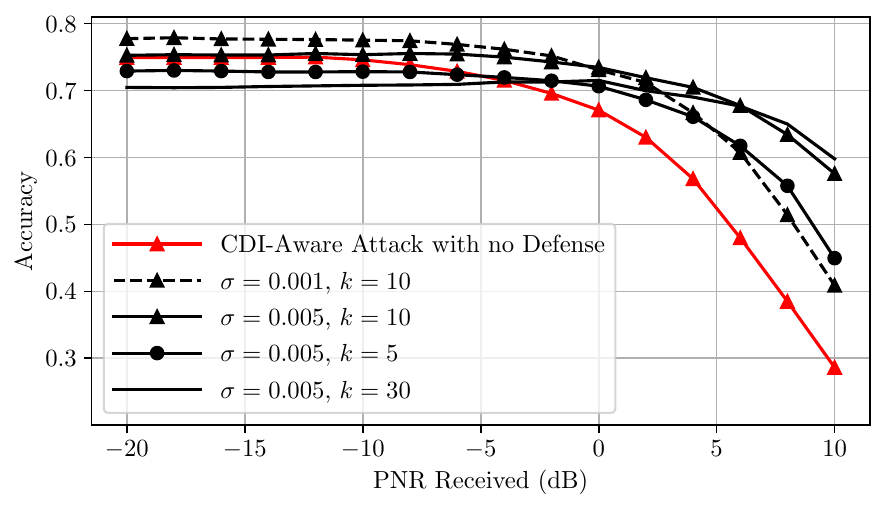}}
    \hfill        
    \caption{Resilience of the target classifier at legitimate receiver against CDI-aware attacks with Gaussian Smoothing for different values of $k$ and $\sigma$}
    \label{fig: gaussian_smoothing}
    \vspace{-12pt}
\end{figure}

\section{CDI-Aware GAN as a Defender}
In this section, we conduct a thorough evaluation of the CDI-aware GAN's effectiveness as a defensive mechanism. Initially, we evaluate a Gaussian smoothing defense strategy. Following this, we proceed to employ adversarial training utilizing our CDI-Aware GAN.

\subsection{Gaussian Smoothing}
In the Gaussian smoothing defense, we fine-tune two critical parameters: the standard deviation of Gaussian noise, $\sigma$, and the number of noisy samples, $k$, added to each training instance. 

Figure~\ref{fig: gaussian_smoothing} presents the impact of Gaussian smoothing on the legitimate receiver's defense. Notable, for PNR $\leq 0$ dB, the classifier effectively counters CDI-aware adversarial attacks. This is particularly evident at $\sigma=0.005$ with $k=\{10, 30\}$, where Gaussian smoothing enhances classifier performance across all PNR levels. 

However, a significant drawback of Gaussian smoothing lies in the need to estimate parameters, $\sigma$ and $k$, which presupposes knowledge of the instantaneous channel between the adversary and legitimate receiver. Accurately determining this channel response and the optimal values for $\sigma$ and $k$ is generally impractical. Consequently, while Gaussian smoothing effectively mitigates adversarial attacks, its feasibility for practical implementation and real-world deployment is substantially constrained. 

\subsection{Adversarial Training}
In this section, we focus on the robustness of the legitimate classifier when adversarially trained with perturbations generated by the proposed CDI-aware GAN. For a comprehensive comparison, we also implement adversarial training defenses based on FGM and PGM. It's crucial to note that the FGM-based adversarial training employs no-channel FGM perturbations, leveraging the receiver's access to both the input to the classifier and the channel between the transmitter and receiver, thereby enhancing the defense. Additionally, we analyze defenses against the MRPP attacks based on both FGM and PGM, introduced in the previous section.

\begin{figure}[t]
    \centering
    {\includegraphics[width=\linewidth]{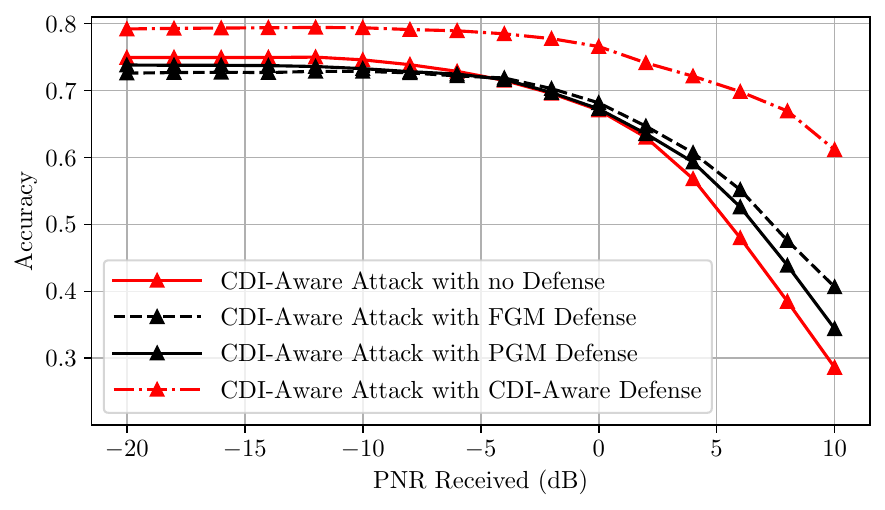}}
    \caption{Our proposed CDI-aware attack against different adversarial training defenses, namely, FGM defense, PGM defense, and CDI-aware GAN defense.}
    \label{fig: adversarial_training}
    \vspace{-12pt}
\end{figure}

Figure~\ref{fig: adversarial_training} illustrates the accuracy of classifiers adversarially trained against our CDI-aware attack. We observe that defenses employing FGM and PGM do not significantly enhance the robustness of legitimate classifiers against our CDI-aware attack. In contrast, adversarial training with the CDI-aware GAN not only counteracts CDI-aware adversarial attacks but also markedly improves the classifier's accuracy.

\begin{figure}[t]
    \centering
    {\includegraphics[width=\linewidth]{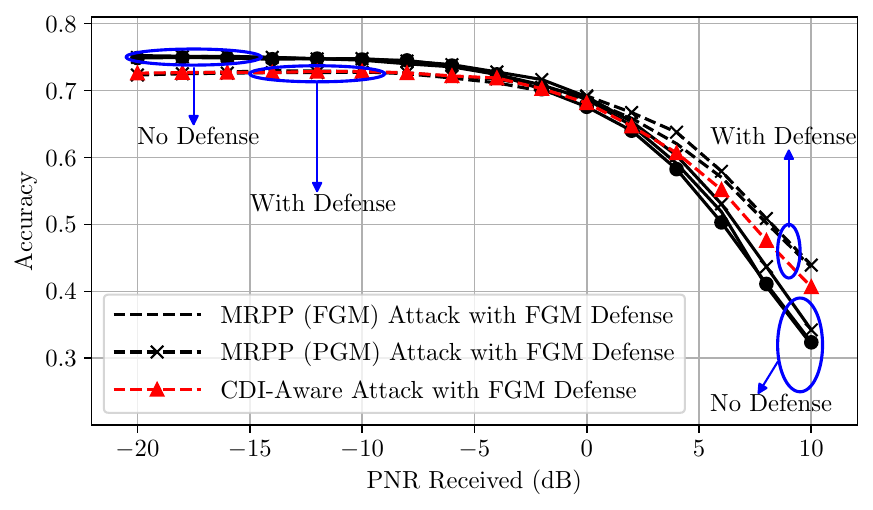}} 
    \caption{FGM-based adversarial training defense against different attacks. Solid lines represent results under no defense and dashed lines with defense.}
    \label{fig: fgmdefender}
\end{figure}

\begin{figure}[t]
    \centering
    {\includegraphics[width=\linewidth]{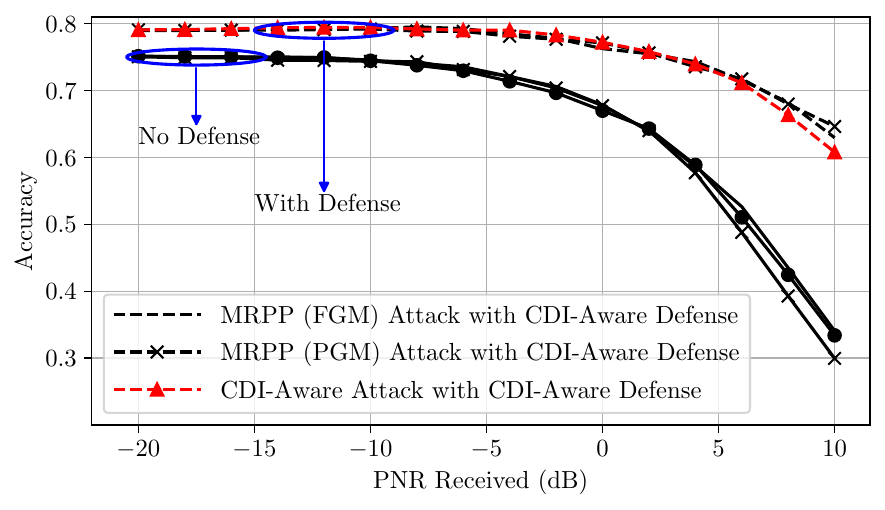}}
    \caption{Our proposed CDI-aware adversarial training defense against different attacks. Solid lines represent results under no defense and dashed lines with defense.}
    \label{fig: ourdefender}
    \vspace{-12pt}
\end{figure}

Figs.~\ref{fig: fgmdefender} presents the resilience of the FGM-based adversarial training defense. The FGM-based defense, especially against our proposed CDI-aware attack, appears less effective. However, as depicted in Figure~\ref{fig: ourdefender}, our findings indicate that adversarial training using perturbations from the CDI-aware GAN substantially increases the classifier's robustness. The CDI-aware GAN, functioning as a defender, successfully neutralizes various adversarial attacks while enhancing the classifier's overall accuracy. This is attributed to the proposed generator model's ability to learn a distribution that not only encompasses channel distribution information but also shares characteristics with FGM perturbations, all while remaining indiscernible from additive white Gaussian noise. As a result, a classifier trained adversarially with perturbations from the CDI-aware GAN emerges as a robust system, proficient in accurately classifying a wide range of adversarial examples.

\section{Experiment Set-up}
In our experimental results, we utilize the GNU radio ML dataset RML2016.10a~\cite{o2016radio}, focusing our analysis specifically on samples with an SNR of $10$ dB. The classifier employed at the legitimate receiver is VTCNN2~\cite{o2017introduction}. We model the propagation channel between the adversary and the legitimate receiver as Rayleigh fading combined with shadowing and path loss, in line with the channel models compared in ~\cite{Sinha2023}. 

We train the CDI-aware GAN with $\alpha = 1$ and $\beta = 50$. The generator, $G$, trains using the Adam optimizer with a $10^{-3}$ learning rate. The discriminators $D_1$ and $D_2$ utilize the Adam optimizer with a learning rate of $10^{-6}$ and employ the \emph{f1 score} as the evaluation metric. Both discriminators are trained to achieve an \emph{f1 score} of $0.5$. 

It is important to note that the CDI-aware generator model architectures for the attacker and defender in our experiments are distinct; they do not share model details with each other. 

\section{Conclusion}
This paper has presented a comprehensive study on the application of a CDI-aware GAN in the context of wireless communication security, both as an attacker and a defender. The experimental results have demonstrated the effectiveness of the CDI-aware GAN in crafting adversarial perturbations that are robust to channel effects, outperforming traditional approaches like FGM and PGM-based MRPP strategies by 3 dB. 

In defense mode, the CDI-aware GAN proved to be a formidable tool. While Gaussian smoothing and FGM-based adversarial training offered some levels of protection, they were limited either by practical implementation challenges or by effectiveness against sophisticated attacks. In contrast, the CDI-aware GAN, through adversarial training, enhanced the classifier's robustness, effectively neutralizing various adversarial attacks and improving overall classification accuracy.

In conclusion, the CDI-aware GAN represents a significant advancement in the field of wireless communication security. Future work could explore the analysis of different channel uncertainty levels and surrogate classifiers.

\end{document}